# Modular Inflatable Composites for Space Telescopes

Aman Chandra
Space and Terrestrial Robotic
Exploration (SpaceTREx) Laboratory
Dept. of Aerospace and Mechanical Eng.
University of Arizona
achandra@email.arizona.edu

Jekan Thangavelautham
Space and Terrestrial Robotic
Exploration (SpaceTREx) Laboratory
Dept. of Aerospace and Mechanical Eng.
University of Arizona
jekan@email.arizona.edu

*Abstract*—There is an every-growing need to construct large space telescopes and structures for observation of exo-planets, main-belt asteroids and NEOs. Space observation capabilities can significant enhanced by large-aperture structures. Structures extending to several meters in size could potentially revolutionize observation enabling technologies. These include star-shades for imaging distant objects such as exo-planets and high-resolution large aperture telescopes. In addition to size, such structures require controllable precision surfaces and high packing efficiencies. A promising approach to achieving high compaction for large surface areas is by incorporating compliant materials or gossamers. Gossamer structures on their own do not meet stiffness requirements for controlled deployment. Supporting stiffening mechanisms are required to fully realize their structural potential. The accuracy of the 'active' surface constructed out of a gossamer additionally also depends on the load bearing structure that supports it. This paper investigates structural assemblies constructed from modular inflatable membranes stiffened pneumatically using inflation gas. These units assembled into composites can yield desirable characteristics. We present the design of large assemblies of these modular elements. Our work focuses on separate assembly strategies optimized for two broad applications. The first class of structures require efficient load bearing and distribution. Such structures do not need high precision surfaces but the ability to efficiently and reliably transmit large loads. This can be achieved using a hierarchical assembly of inflatable units. They also need to be stiffer as a collective assembly as compared to their constituent modules. Preferential placement of varying modular units leads to local stiffness modulation. This in-turn helps modify load transmission characteristics. Applications of such structures also extend to deployable drag-chutes or aero-breaking devices for atmospheric maneuvering. The second are structures with precision surfaces for optical imaging and high-gain communication apertures. We demonstrate over-constrained modular assemblies exhibiting elastic averaging when assembled with a very large number of modules. Averaging effects are amplified with the number of sub-units approaching required surface precision with a large enough number. Our work includes fundamental structural studies to evolve feasible sizing schemes for both classes of structures. A structural analysis strategy using discrete finite elements has been developed to simulate the assembled behavior of modular units. The structural model of each inflatable unit has been extended from our previous work to approximate each unit as a 3-dimensional truss system. Analysis results are compared with full scale simulations on commercial analysis package LS-Dyna. Our analysis leads to an understanding of the extent to which inflatables can be scaled up effectively. Critical geometric design considerations are identified for stowed and deployed states of each structure. We propose designs of compliant hinges between structure to assemble even large units. Further work includes prototype development and deployment force measurement to validate the structural model.

TABLE OF CONTENTS

## 1. INTRODUCTION

Enhancements in deep space imaging techniques have led to several new discoveries. Imaging bodies such as exo-planets and near-earth objects (NEO's) with increasing precision is a key enabling technology. Reflector mirrors with increasingly large apertures are critical towards achieving such capabilities. The Hubble Space Telescope (HST) consists of the largest primary mirror currently on-orbit with a diameter of 2.4 meters. Current research is focused on increasing aperture sizes to several meters in diameter. The James Webb Space Telescope with a primary mirror spanning about 6.5 meters in diameter is currently under development (Figure 1).

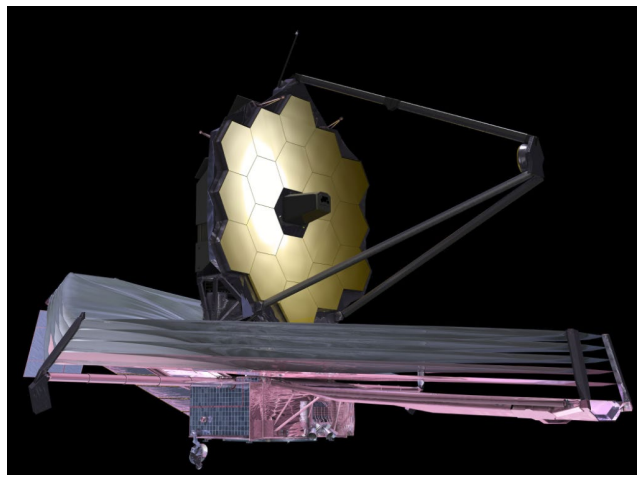

**Figure 1. James Webb Space Telescope**

JWST's large apertures offer several advantages. Collecting energy from distant bodies requires apertures in excess of 10

meters [1]. Large sizes also help enhance diffraction limited angular resolution which is a function improves with an increase in size. In addition to imaging of deep space objects, there is also an increasing requirement to enhance space situational awareness and earth observation. The ability to place large an optical mirror several meters in size in the GEO observing Earth can provide unprecedented coverage time and resolution [2].

Scaling up the size of space-based mirrors poses formidable challenges. These range from high precision manufacturing methods, precision deployment mechanisms, mass and volume restrictions. The size of the largest structure than can be placed in the launch shroud of state of the art space launch systems is currently limited to around 8 meters in diameter [3]. The requirement for mirrors with much larger apertures will need techniques to efficiently package, store and deploy these mirrors without affecting surface precision.

Among deployable mirror technologies, research focus has moved towards optical technology such as segmented mirrors and trans-missive diffractive optical elements [4]. While these methods hold promise, they exhibit significant structural complexities thus adding costs and reducing launch reliability. To realize optical surfaces in excess of tens of meters in diameter, a more efficient mechanism is needed that begins to solve some of these problems. Table 1 shows a comparison between various deployable mirror technologies.

**Table 1. Deployable mirror technologies**

| Mirror technology | Packing Efficiency | Areal Density | Scalability |
|---|---|---|---|
| Fixed Mirrors | 1:1 | 1-2 kg | Low |
| Linkage Systems | 5:1 | 1-2 kg | Medium |
| Membranes | 20:1 | 0.3 – 0.5 kg | High |

As can be observed membrane technology holds greatest potential in terms of areal density and potential to scale up. The areal density or mass per unit surface are is also the least in the case of membrane mirrors. Conventionally, membrane mirror are defines as having small enough thicknesses that stretching or tensile stresses dominate over bending response.

There are, however several challenges associated with membrane technology. For most optical or infrared wavelength applications, the performance of the mirror is estimated in terms of spatial and temporal perturbations of the generated wave-fronts at the wavelength of interest. Tolerance towards these perturbations is dependent on the specific application in question. In the case of optical and infrared bands, the RMS perturbation allowance stands at around 0.4-4 μm. [5]. These requirements translate into an allowable RMS shape error of about one hundredth of the wavelength of imaging. In addition, dynamic vibration modes need to be restricted to controllable levels.

A combination of strategies are required to enable membrane reflectors. These range from structural design, optical design, manufacturing technology and spacecraft control. Our present paper looks into structural design strategies for membrane reflectors. Our present work extends our previous work on modular inflatable space structures [6]. This work investigates pneumatically inflated membranes as structural units of large assemblies. We demonstrate a basic analysis strategy that attempts to characterize precision obtained by such assemblies. We look into strategies of enhancing precision by introducing structural constraints. Performance of such assemblies is also studied from a load bearing point of view. Apart from the reflector itself, supporting trusses and structures are of importance as they affect structural behavior. We also look at assembly topologies based on membrane units of varying geometries.

## 2. RELATED WORK

Membrane structures for space have been investigated since the 1950's as an alternative to conventional structures [7]. The ECHO balloon project by NASA in the 1960's [8] was the first successful demonstration of inflatable membranes on a large scale. ECHO 1 and 2 were large orbiting spheres constructed out of metallized Mylar and spanned over 30 meters in diameter. They operated as reflectors for radio waves in the frequency range of 162 to 2390 Mhz and successfully stayed in orbit for several years. Their success triggered multiple investigations into membrane reflectors spanning several meters in diameter. The next successful attempt was that of a 16-meter diameter reflector deployed onboard STS-77 in the late 1990's. The experiment was termed as the 'inflatable antenna experiment' and demonstrated the on-orbit deployment of an inflatable membrane reflector [9].

Membrane mirror technology has been researched extensively since the 1990's. Initial architectures employed stretched flat membrane mirrors. Stretched membrane with electrostatically induced curvature [10] were studied with varying success. More stable architectures with low areal densities such as lenticular inflatables have been developed by L'GARDE and SRS technologies [2]. These mainly consisted of a primary membrane mirror supported by lenticular inflatable panels. Primary challenges observed were the persistence of uncontrolled frequency modes and lack of adequate shape control.

Apart from membrane mirror technology itself, advancements in associated optical technologies have also received considerable attention in the past few years. Ball Aerospace's and DARPA's MOIRE program [4] aims to place a GEO based 20-m diameter operational telescope for the persistent monitoring of the ground. Advances in transmissive diffractive optics techniques have greatly relaxed surface precision requirements from telescope mirrors. Figure 2 shows a conceptual image of the MOIRE telescope showing segmented elements.

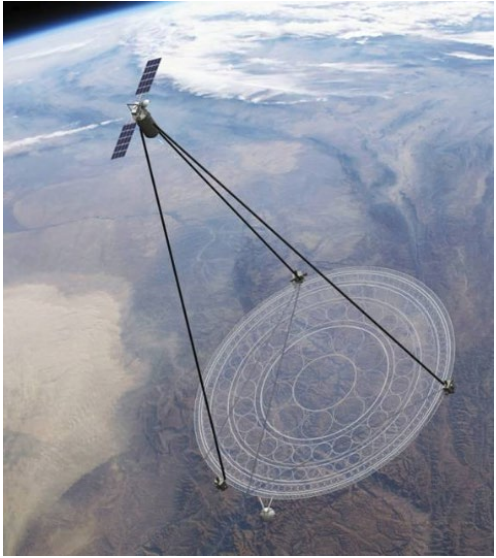

**Figure 2. MOIRE Space Telescope [4]**

Efficient imaging of distant objects such as exoplanets requires large scale occultors to work in conjunction with telescopes. Once such concept as described in NASA's 2020 decadal survey [11] is NASA's Habitable Exoplanet Observatory (HabEx) shown in Figure 3. The mission is designed to observe exoplanets with the potential of sustaining life. It consists of a deployable occultor termed as 'Startshade.' The Starshade is envisaged to be a 72 m in diameter membrane-based tensegrity structure which deploys from a conventional launch vehicle. Occultors to suppress light from stars popularly also known as star-shades are a similar class of structures being developed. The design consists of a deployable truss structure that acts as a primary support imparting the membrane with stiffness [12]. A network of spokes is used to effect intermediate control on the structure. The Starshade concept highlights the scale and complexity involved in the deployment of space structures of that size.

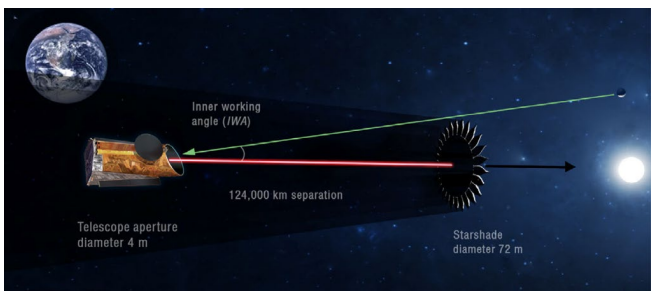

**Figure 3. NASA's Habitable Exoplanet Observatory (HabEx) [11].**

Among surveyed membrane mirror technologies, it becomes evident that major technology areas need further development before a reliable membrane reflector can operate on-orbit. From an optical standpoint, active wave-front control for manipulating shape of the optical signal would be necessary to bring down mirror surface accuracy requirements to within tolerance. From the structural standpoint, design emphasis must extend to include packaging and deployment geometries in addition to final desired shapes. Active sensing and control of the membranes are necessary to achieve long term reliability.

Space telescopes also contain peripheral supporting structures. They play an instrumental role in precise placement and alignment of membrane optics. Besides providing structural stability to the deployed mirrors, they also contribute towards vibration suppressions and modulation of natural frequency modes of the structure as a whole. Dorsey and Mikulas [13] discuss basic design considerations for truss and beam based structural system design. Due to the high precision required for imaging applications, major research has focused on rigid linkages and high-stiffness mechanisms. The field of compliant mechanisms and soft-robotic manipulators has received considerably less attention from the space community. While precision and accuracy of placement are necessary, it is also important for the mirror's support structure to support compact stowage and reliable deployment.

## 3. METHODOLOGY

Compliant gossamer structures such as membranes cannot resist compressive loads. This gives rise to local buckling phenomena and wrinkle formation. Investigations into wrinkling phenomenon [14] show that wrinkle ridges transmit loads at such locations making them susceptible to stress concentrations. To effectively capture the benefits of a gossamer membrane based structure, a reliable structural system needs to be developed. Our strategy lies in replacing complex large membranes with smaller membrane units. The purpose of adopting such a strategy is two-fold. The first is to understand structural behavior at an elemental level. This would be useful in developing localized membrane control strategies. The second is to constrain boundary conditions on membrane units. This would aid to restrict highly non-linear and uncontrolled structural modes. In our investigation of membrane unit assemblies, we look into their response from two perspectives. One is from that of precision of shape. To be feasible optical membranes, the ability of a structural system to retain a specified shape with precision would be necessary. Though the shape of membrane optics would depend on the specific application, for the purpose of our analysis we consider a parabolic reflector. Strategies for over-constraining such assemblies is discussed that could potentially average out surface errors once sufficient scales are achieved. Other than shape precision, we also look into load bearing abilities of such membranes. Assembly techniques based on hierarchical topologies are discussed and their advantages pointed out.

*Inflatable membrane beam elements*

We build up on our previous work [6, 20-22] on inflatable beam elements which have been chosen as the structural unit for our analysis.

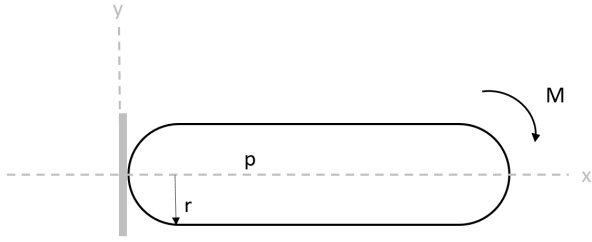

**Figure 4. Inflatable Assembly Abstraction**

Figure 4 shows an inflatable beam element comprising of a cylindrical membrane of radius r inflated with internal pressure p having thickness. Six degrees of freedom include three translation axes and three rotation axes. For an inflatable beam element, maximum loads in terms of bending moments are given by equation (1) [15]

$$\frac{|M_c|}{pr^3} = \left(\frac{1-\upsilon-3\upsilon^2}{1-\upsilon-\upsilon^2}\right)\pi \qquad (1)$$

Here $M_c$ is the moment about the cylinder's centroid. In terms of curvature, the governing equation is a function of buckling angle $\theta_o$

$$\frac{d^2y}{dx^2} = \frac{M}{Etr^3(\pi - \theta_o + \sin(\theta_o)\cos(\theta_o))} \qquad (2)$$

To study the structural behavior as an assembly we construct Euler-Bernoulli beam networks of individual membrane elements.

*Euler-Bernoulli beam networks*

The numerical simulation of large continua assemblies is computationally expensive and complex. It has been shown that in the case where it is physically possible to break down the continuum into discrete well characterized blocks, discrete assemblies of such blocks can yield fairly accurate results at much lower cost [16]. We apply a similar principle to the analysis of discrete membrane unit assembles. In essence, each inflatable unit acts as a physical finite element [17]. Figure 5 shows an abstraction of the modelled beam network

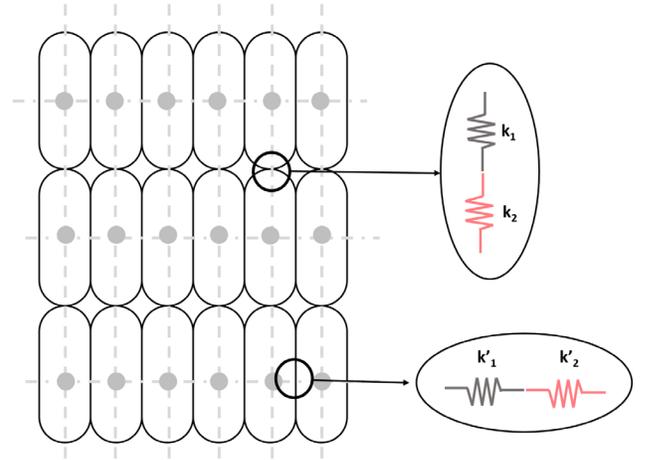

**Figure 5. Inflatable Assembly Abstraction**

Each membrane unit represented at each node takes on an effective elastic modulus k. Since the behavior of such members is not symmetric, their elastic modulus has directionality. As shown in Figure 5, longitudinal and lateral contacts between separate inflatable units are modelled as having an average elastic modulus as follows:

$$k_c = \frac{2k_1 k_2}{k_1 + k_2} \qquad (3)$$

A finite element formulation based on the above assumption is used to find the structure's stiffness matrix. Since the membrane stiffness is modelled on a beam model, each node represents 6 degrees of freedom. The elemental stiffness matrix can be shown as:

$$\begin{pmatrix} F_x \\ F_y \\ F_z \\ M_{\theta_x} \\ M_{\theta_y} \\ M_{\theta_z} \end{pmatrix} = [k] \begin{pmatrix} X_1 \\ X_2 \\ X_3 \\ \theta_{x_1} \\ \theta_{y_1} \\ \theta_{z_1} \end{pmatrix} \qquad (4)$$

*Elastic averaging for structural precision*

Elastic averaging has been explored for high precision discrete assemblies [18]. It is based on the principle of over-constraining a unit in an assembly. As opposed to a rigid linkage system, where minimal constraints imposed on each member, in our analysis we over-constrain the assembly. Additionally, each constraint or contact is compliant in nature as it is defined by its elasticity and not rigid body kinematics. Careful design of compliant contacts leads to a solution where positional inaccuracies of each member unit average out and are minimized once an assembly of sufficient scale is developed. We look to apply the same principle on a doubly curved assembly structure as shown in Figure 6.

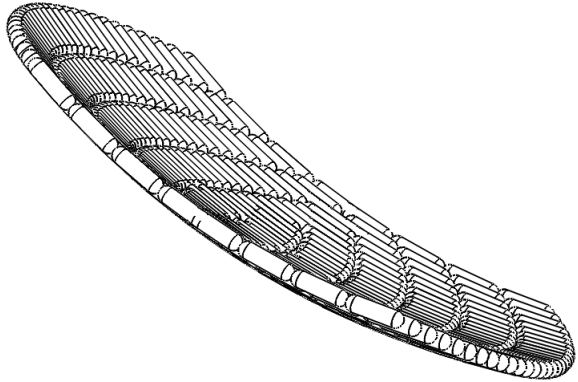

**Figure 6. Double curve inflatable assembly.**

In such an assembly, we impose shape precision requirement and solve to find out optimal contact elasticities. We repeat the calculation over several cases to check if the inaccuracies in shape begin to decline with an increase of assembled components.

*Assembly topologies*

In addition to shape precision, we are also interested in understanding the load bearing requirements of inflatable unit assemblies. Discrete hierarchical assemblies [19] for 3D beam elements have been investigated. We study the load transfer characteristics of membrane units with varying geometry. The objective is to evolve inflatable unit network topologies that can be optimized for strength and surface precision. Candidate membrane element geometries include cylindrical beams and toroidal shells as shown in Figure 7. We conduct a full scale finite element analysis using commercial software package LS-Dyna.

**Table 2. Analysis model properties**

| Analysis Model Property | Value |
|---|---|
| Young's modulus | 4.89 x $10^9$ Pa |
| Poisson ratio | 0.38 |
| Density | 1380 kg/$m^3$ |
| Thickness | 2.54 x $10^{-6}$ m |
| Internal pressure | 6.89 Pa |

Structural simulations are conducted to analyze stress generation over pneumatic deployment time periods. Our material choice for analysis is Mylar membrane. Table 2 lists important material and analysis model properties that we used. The choice of these geometries was primarily due to ease of manufacture and stable deployment characteristics.

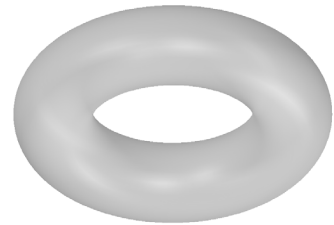

(a) Toroidal Unit

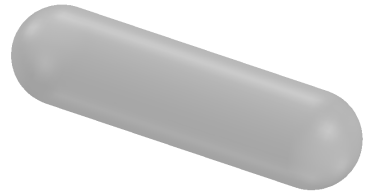

(a) Beam Unit

**Figure 7. Candidate inflatable elements.**

# 4. RESULTS AND DISCUSSION

*Compliant gossamer structures*

Figure 8 shows displacements on individual membrane elements and with longitudinal contact conditions. The displacements plotted are developed over time corresponding to inflation to the final equilibrium pressure. It can be observed that for the same loading a modulation in elastic contacts can cause a large deflection as compared to the case without contacts. Further, we note that the final deflection can be modulated by varying the contact stiffness values. Maximum beam deflection in this case is linear and attains a maximum average of 0.33m.

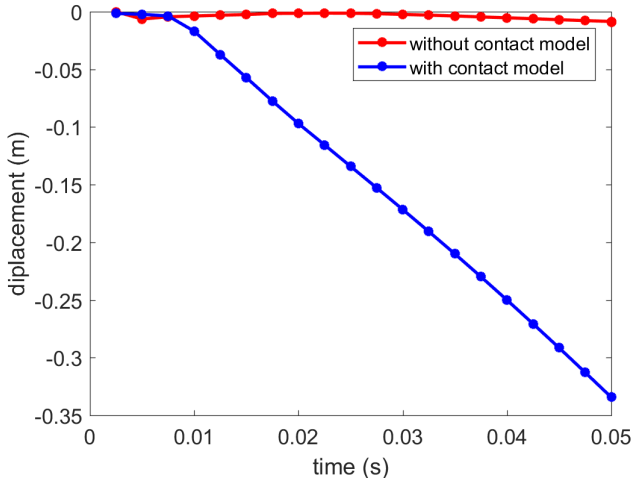

**Figure 8. Beam deflection behavior.**

The magnitude of each linear deflection can be modulated based on the contact elasticity at the preceding and succeeding ends of each membrane unit. Hence, by choosing an adequate elasticity modulating scheme, a surface of desired curvature can be constructed using a sufficiently large number of membrane unit. Figure 9 shows a similar comparison between each case in terms of equivalent stress generation.

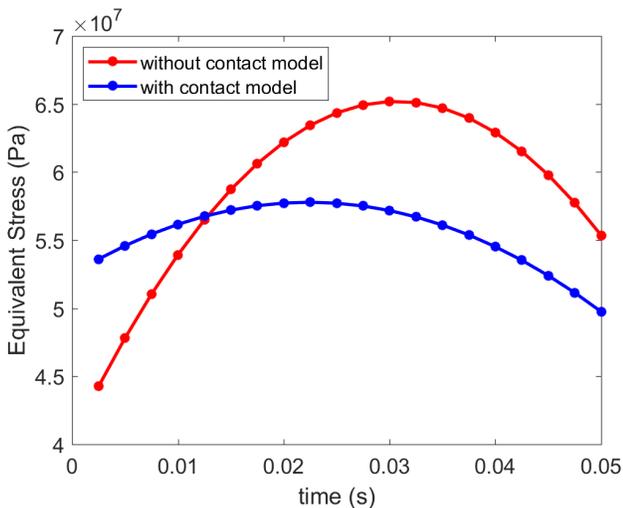

**Figure 9. Equivalent stress distribution.**

It can be observed that for a larger deflection, generated stresses are reduced in the case of compliant contacts. This is consistent with the assumption that elastic contacts cause a more uniform distribution of stresses over the membrane thereby reducing local buckling tendencies.

*Assembly topologies*

Figure 10 shows a comparison between inflatable membrane geometries in terms of equivalent stress generated over time for identical load cases. It can be seen that stresses in the case of beam units grow at a much faster rate as compared to toroidal elements. The *x*-axis represents the time taken to inflate each individual membrane element to the specified equilibrium pressure.

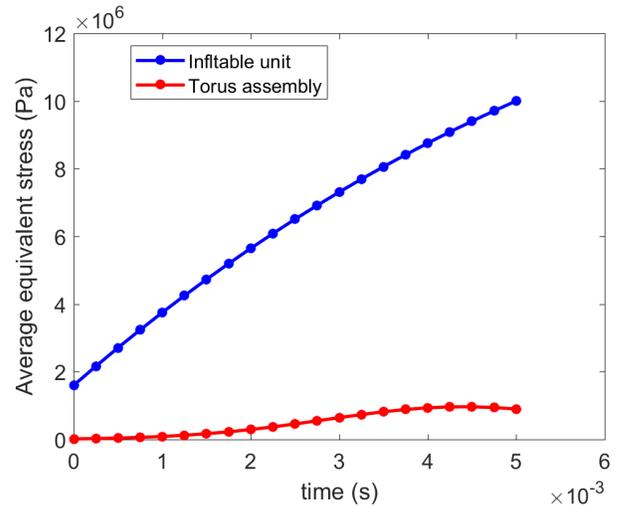

**Figure 10. Equivalent stress distribution.**

Toroidal elements on the other hand show steady stress states with a much lower equivalent stress magnitude. This points towards a much more uniform distribution of loads over the toroidal circumference as opposed to beam element which see sharp transitions.

From the above analysis it can be said that inflatable elements with multiple geometries can be employed to form a composite structure. While beam elements exhibit less uniform load distribution, their deflective capabilities can be used for components of the structure where specified shapes are desired. Toroidal elements on the other hand show enhanced load transfer abilities and can be positioned in regions of the structure more suited towards a load transfer role.

# 5. Conclusions and Future Work

In this paper we develop a methodology towards designing large space structures built as assemblies of discrete inflatable membrane elements. A first order model is developed to incorporate compliant contacts between each inflatable unit. The model is evaluated from the point of view of two major applications. In the first application, we intend to maximize precision of assembled shape and in the second, maximize load bearing abilities of such assemblies.

We show that modulating elastic contacts can be used to obtain varying rates of deflection. For large assemblies, this would translate to modification of local curvatures which could in-turn be used for applications such as membrane reflectors and reflector antennas. A full scale finite element simulation is performed on different membrane unit geometries. In the context of load bearing, we find that toroidal elements while well suited for load bearing applications, can be used to effect adequate shape control. On the other hand, beam elements have inferior load transfer abilities but can be structurally tuned to take on a specific shape. Through this work we aim to develop a structural design and analysis framework that can be used over several load cases and for varying geometries.

Future work includes enhancing the framework's fidelity by incorporated directional stiffness effects. This would lead to an optimization problem, where the solution would yield contact elasticity patterns to achieve local stiffness and shape modulation. This would be followed by full scale designing and prototyping of contact mechanisms to create inflatable membrane assemblies.

## Biography

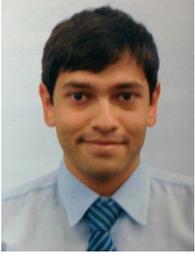

***Aman Chandra*** *received an MS in Aerospace Engineering at Arizona State University. He is currently a PhD student at the University of Arizona department of Aerospace and Mechanical Engineering. His master's thesis dissertation is on the design and optimization of gossamer space structures for small satellites. His research interests include structural dynamics, computational geometry and multidisciplinary design optimization.*

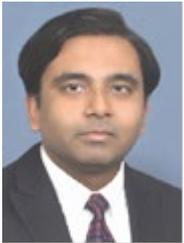

***Jekanthan Thangavelautham*** *has a background in aerospace engineering from the University of Toronto. He worked on Canadarm, Canadarm 2 and the DARPA Orbital Express missions at MDA Space Missions. Jekan obtained his Ph.D. in space robotics at the University of Toronto Institute for Aerospace Studies (UTIAS) and did his postdoctoral training at MIT's Field and Space Robotics Laboratory (FSRL). Jekan is an assistant professor and heads the Space and Terrestrial Robotic Exploration (SpaceTREx) Laboratory at the University of Arizona. He is the Engineering Principal Investigator on the AOSAT I CubeSat Centrifuge mission and is a Co-Investigator on SWIMSat, an Airforce CubeSat mission concept to monitor space threats.*